\documentclass{article}
\usepackage{amsfonts,amssymb,amsmath,amsthm,graphicx}
\usepackage[english]{babel}
  \newtheorem{lem}{Lemma}
  \newtheorem{thm}{Theorem}
  \newtheorem{cor}{Corollary}

\let\phi=\varphi

\let\theta=\vartheta
\newcommand{\eps}{{\varepsilon}}

\newcommand{\bC}{{\mathbb C}}

\newcommand{\bP}{{\mathbb P}}

\newcommand{\bZ}{{\mathbb Z}}

\DeclareMathOperator{\GL}{GL}

\newcommand{\ket}{\rangle}

\begin{document}
\title{\Large Projective Ring Line of an Arbitrary\\ Single Qudit\\~}

\author{Hans Havlicek$^{1}$ and Metod Saniga$^{2}$\\
\\
\normalsize $^{1}$Institut f\" ur Diskrete Mathematik und Geometrie\\
\normalsize  Technische Universit\" at Wien, Wiedner Hauptstrasse 8-10\\
\normalsize A-1040 Vienna, Austria\\
\normalsize  (havlicek@geometrie.tuwien.ac.at)
\\ \\
\normalsize  $^{2}$Astronomical Institute, Slovak Academy of Sciences\\
\normalsize  SK-05960 Tatransk\' a Lomnica, Slovak Republic\\
\normalsize  (msaniga@astro.sk)}

\date{\small (December 27, 2007)}

\maketitle

\begin{abstract}\noindent
As a continuation of our previous work (arXiv:0708.4333) an algebraic geometrical
study of a single $d$-dimensional qudit is made, with $d$ being {\it any} positive integer. The study is based on an intricate relation
between the symplectic module of the generalized Pauli group of the qudit and the fine structure of the projective line over the (modular)
ring $\bZ_{d}$. Explicit formulae are given for both the number of generalized Pauli operators commuting with a given one and the number
of points of the projective line  containing the corresponding vector of $\bZ^{2}_{d}$. We find, remarkably, that a perp-set is not a
set-theoretic union of the corresponding points of the associated projective line unless $d$ is a product of distinct primes.
The operators are also seen to be structured into disjoint `layers' according to the degree of their representing vectors. A brief comparison
with some multiple-qudit cases is made.\\

                  Algebra of Generalized Pauli Operators

\end{abstract}

\vspace*{.5cm}
\section{Introduction}
In our recent paper \cite{hs} we introduced a general algebraic geometrical framework underlying the structure of the
generalized Pauli group associated with a specific single $d$-dimensional qudit. The backbone of this framework is the bijection between sets of operators/matrices of the group and vectors over the modular ring $\bZ_{d}$. This bijection
enabled us, for $d$ being a product of distinct primes, to completely rephrase the group's commutation algebra in terms of the structure of and interplay between free cyclic submodules of $\bZ_{d}^{2}$ {\it aka} points of the projective line
defined over $\bZ_{d}$. In this paper we shall tackle the general case (i.\,e., $d$ being any positive integer), making thus
the treatment of a {\it single} qudit complete.

\section{Single $d$-qudit, its qeneralized Pauli group,\\ symplectic module and projective ring line}
In this section we simply set up the notation and recollect some basic technical results from our previous paper \cite{hs}
to be needed in the sequel.
\par
Let $d>1$ be an integer and $\bZ_d:=\{0,1,\ldots,d-1\}$. Addition and
multiplication of elements from $\bZ_d$ will always be understood modulo $d$.
We consider the $d$-dimensional complex Hilbert space $\bC^d$ and denote by
\begin{equation*}
   \{\, |s\ket : s\in\bZ_d \}
\end{equation*}
a computational basis of $\bC^d$. Taking $\omega$ to be a fixed
primitive $d$-th root of unity (e.\,g.,\ $\omega=\exp(2\pi i/d)$), we define
unitary $X$ (``shift") and $Z$ (``clock") operators on $\bC^d$ via $X|s\ket =
|s+1\ket$ and $Z|s\ket = \omega^s |s\ket$ for all $s\in\bZ_d$.
With respect to our computational basis the matrices of $X$ and $Z$ are
\begin{equation*}
    \begin{pmatrix}
    0      & 0     &\ldots &0      & 1\\
    1      & 0     &\ldots &0      & 0\\
    0      & 1     &\ldots &0      & 0\\
    \vdots &\vdots &\ddots &\vdots &\vdots\\
    0      & 0     &\ldots &1      & 0
    \end{pmatrix}
    \mbox{~~and~~}
    \begin{pmatrix}
    1      & 0     &0        &\ldots&  0\\
    0      &\omega &0        &\ldots&  0\\
    0      & 0     &\omega^2 &\ldots& 0\\
    \vdots &\vdots &\vdots   &\ddots &\vdots\\
    0  & 0 & 0     &\ldots   &\omega^{d-1}
    \end{pmatrix},
\end{equation*}
respectively.
The (generalized) \emph{Pauli group} generated by $X$ and $Z$ will be
denoted as $G$.
For all $s\in\bZ_d$ we have $XZ|s\ket = \omega^s|s+1\ket$ and $ZX|s\ket =
\omega^{s+1}|s+1\ket$. This gives the basic relation
\begin{equation}\label{eq:xzzx}
    \omega XZ = ZX
\end{equation}
which implies that each element of $G$ can be
written in the unique \emph{normal form}
\begin{equation}\label{eq:normalform}
    \omega^a X^b Z^c \mbox{~~for some integers~~}a,b,c\in \bZ_d.
\end{equation}
From (\ref{eq:xzzx}) it is readily seen that
\begin{equation*}
    (\omega^a X^b Z^c) (\omega^{a'} X^{b'} Z^{c'}) = \omega^{b'c + a+a'}
X^{b+b'} Z^{c+c'},
\end{equation*}
which shows that $G$ is a non-commutative group of order
$d^3$.
Next, the \emph{commutator\/} of two operators $W$ and $W'$ is
\begin{equation}\label{eq:defcommutator}
    [W,W'] := W W'W^{-1}{W'} ^{-1}
\end{equation}
which in our case acquires the form
\begin{equation*}\label{eq:commutator}
    [\omega^a X^b Z^c, \omega^{a'} X^{b'}Z^{c'}] = \omega^{cb'-c'b}I.
\end{equation*}
Recall that two operators commute if, and only if, their commutator (taken in
any order) is equal to $I$ (the identity matrix).
\par
There are two important normal subgroups of $G$:
its \emph{centre\/} $Z(G)$ and its \emph{commutator subgroup} $G'$, the two being identical
\begin{equation}\label{eq:G'}
    G' = Z(G) = \{\omega^a I: a\in\bZ_d \}.
\end{equation}
The bijective mappings
\begin{equation*}
    \psi : \bZ_d \to G' : a \mapsto \omega^a I,
\end{equation*}
\begin{equation*}
    \phi : \bZ_d^2 \to G/G' : (b,c) \mapsto G'X^bZ^c.
\end{equation*}
and their inverses yield a mapping\footnote{Of course the symbol $[\cdot,\cdot]$
has two different meanings in (\ref{eq:defcommutator}) and
(\ref{eq:alternating}).}
\begin{equation}\label{eq:alternating}
    [\cdot,\cdot] : \bZ_d^2 \to\bZ_d : \big((b,c),(b',c')\big) \mapsto cb'-c'b
\end{equation}
which just describes the commutator of two elements of $G$ (given in normal
form) in terms of our $\bZ_d$-module. The
mapping (\ref{eq:alternating}) can be rewritten in the following convenient form
\begin{equation}\label{eq:determinant}
    \big[(b,c),(b',c')\big]
    = (b,c)\begin{pmatrix} 0 &-1\\1&\hphantom{-}0\end{pmatrix}\begin{pmatrix}b'\\c'\end{pmatrix}
    = \det \begin{pmatrix} b' &c'\\b\hphantom{'}&c\hphantom{'}\end{pmatrix}
\end{equation}
which implies that $[\cdot,\cdot]$ is a bilinear alternating form on
$\bZ_d^2$.  As usual, we write $(b,c)\perp(b',c')$ if
$\big[(b,c),(b',c')\big]=0$ and speak of \emph{orthogonal} (or: \emph{perpendicular}) vectors (with
respect to $[\cdot,\cdot]$). As our alternating bilinear form is non-degenerate, we have indeed a \emph{symplectic module}.
\par
The set of operators in $G$ which commute with a fixed
operator $\omega^aX^bZ^c$ corresponds to the \emph{perpendicular set} (shortly
the \emph{perp-set}) of $(b,c)$, viz.\
\begin{equation*}
    (b,c)^\perp := \big\{(u,v)\in\bZ_d^2 : (b,c)\perp (u,v) \big\}.
\end{equation*}
Being closed under addition and multiplication by ring
elements and fulfilling the condition
\begin{equation*}\label{eq:perpset}
    \bZ_d(b,c)\subset(b,c)^\perp,
\end{equation*}
the set $(b,c)^\perp$ is a $\bZ_d$-submodule of $\bZ_d^2$.

A full algebraic geometrical meaning of perp-sets in $\bZ_d^2$ is revealed after introducing the concept of the projective line over the ring $\bZ_d$.  We sketch here only some basic notions
and results, referring the interested reader to \cite{bh}--\cite{blhr} for further details and proofs.
\par
Let us consider any vector $(b,c)\in\bZ_d^2$. It generates the cyclic
submodule
\begin{equation*}
    \bZ_d(b,c) = \{(ub,uc):u\in\bZ_d\}.
\end{equation*}
Such a cyclic submodule is called \emph{free\/}, if the mapping
$u\mapsto(ub,uc)$ is injective. In this case the vector (or: pair) $(b,c)$ is
called \emph{admissible}. Any free cyclic submodule of $\bZ_d^2$ has precisely
$d$ vectors, including the zero-vector. However, not all vectors $\neq(0,0)$ of
a free cyclic submodule need to be admissible.
In a more geometric language,  a free cyclic submodule of $\bZ_d^2$
is called a \emph{point}. The point set
\begin{equation*}\label{eq:projectiveline}
    \bP_1(\bZ_d):=\{\bZ_d(c,d) : (c,d) \mbox{~is admissible} \}
\end{equation*}
is the \emph{projective line\/} over the ring $\bZ_d$. According to this
definition a point is a set of vectors.

\section{A qudit for $d$ a prime power}\label{sect:primepower}

The case of $d$ being a product of distinct primes was dealt with
in \cite{hs} where the interested reader can find all the details; following
the strategy and employing the findings of this paper, we are now in position to successfully tackle the most general case.
\par
In this section, as a necessary intermediate step, we focus our attention to the case of a single qudit for
$d=p^\eps$, where $p$ is a prime and $\eps\geq 1$ an integer.
Even
though we aim at using representatives from $\bZ_d$ rather than arbitrary
integers (to be reduced modulo $d$), it will be very convenient to represent $0\in\bZ_d$ also by
$d=p^\eps\notin\bZ_d$. We remind the reader that exponents in terms like
$p^\alpha$ or $p^{\alpha+\beta}$ are non-negative integers which \emph{must
not} be reduced modulo $d$. Of course, when speaking about cardinalities of
sets, also no reduction modulo $d$ has to be applied.
\par
Each of the sets
\begin{equation*}\label{eq:ideal-chain}
    \bZ_d = \bZ_d\cdot p^0 \supset \bZ_d\cdot p^1 \supset\cdots\supset\bZ_d\cdot
    p^\eps =\{0 \}
\end{equation*}
is an ideal of the ring $(\bZ_d,+,\cdot)$. We infer from
\begin{equation*}\label{eq:ideal}
    \bZ_d\cdot p^\kappa = \{w p^\kappa : w=1,2,\ldots, p^{\eps-\kappa} \}
\end{equation*}
that $|\bZ_d\cdot p^\kappa|= p^{\eps-\kappa}$ for any
$\kappa\in\{0,1,\ldots,\eps\}$.
\par An element
of $\bZ_d$ has a multiplicative inverse if, and only if, it belongs to $\bZ_d
\setminus \bZ_d\cdot p$. So the elements of $\bZ_d$ without an inverse (i.~e.\
the zero-divisors of $\bZ_d$) are precisely the $p^{\eps-1}$ elements of
$\bZ_d\cdot p$. We note in passing that $\bZ_d\cdot p$ is the only maximal
ideal of the ring $\bZ_d$. So $\bZ_d$ is a \emph{local ring}.
\par
Each element $a\in\bZ_d$ admits a factorisation of the form
\begin{equation}\label{eq:power-rep}
    a = up^\alpha
    \mbox{~~with~~}
    u\in\bZ_d\setminus\bZ_d\cdot p
    \mbox{~~and~~} \alpha\in\{0,1,\ldots,\eps \}.
\end{equation}
Indeed, $a=0$ can be written as $a=1p^\eps$, for $a=1$ holds $a=1p^0$, and for
any other element of $\bZ_d$ the usual decomposition of $a$ into a product of
primes, which uses the arithmetics over $\bZ$, gives also a solution in
$\bZ_d$. The integer $\alpha$ is determined uniquely: It is the smallest
element $\kappa\in\{0,1,\ldots,\eps\}$ such that $a\in\bZ_d\cdot p^\kappa$.
This uniqueness need not hold for $u$. In the case $a=0$ the element $u$ may be
any invertible element of $\bZ_d$. For any $a\neq 0$ the element $u$ is given
up to an additive constant belonging to
\begin{equation*}\label{eq:annihilator}
    \bZ_d\cdot p^{\eps-\alpha} = \{wp^{\eps-\alpha}:w=1,2,\ldots,p^\alpha \}.
\end{equation*}
This set is the \emph{annihilator\/} of $p^\alpha$, i.~e. the set of all
$x\in\bZ_d$ with the property $p^\alpha x=0$.
\par
Let $(b,c)$ be a vector of the $\bZ_d$-module $\bZ_d^2$, where $b=vp^\beta$ and
$c=wp^\gamma$ are factorisations as in (\ref{eq:power-rep}). Then
$\min\{\beta,\gamma\}$ will be called the \emph{degree\/} of $(b,c)$. So this
degree equals the smallest index $\kappa\in\{0,1,\ldots,\eps\}$ such that
$b,c\in\bZ_d\cdot p^\kappa$. It is an easy exercise to show that $(b,c)$ has
degree $\kappa$ if, and only if, the ideal of $\bZ_d$ generated by $\{b,c\}$
equals $\bZ_d\cdot p^\kappa$.

\begin{lem}\label{lem:transitive}
Let $(b,c)$ be a vector of $\bZ_d^2$ with degree $\delta$. Then the following
assertions hold:
\begin{enumerate}
\item
If $A$ is an invertible $2\times2$ matrix over $\bZ_d$ or, in symbols
$A\in\GL_2(\bZ_d)$, then $(b,c)A$ is also a vector of degree $\delta$.
\item
There exists a matrix $M\in\GL_2(\bZ_d)$ such that $(b,c)M=(p^\delta,0)$.
\end{enumerate}
\end{lem}
\begin{proof}
Suppose that $b=vp^\beta$ and $c=wp^\gamma$ are factorised according to
(\ref{eq:power-rep}). Given a matrix $A=(a_{jk})\in\GL_2(\bZ_d)$ we obtain from
\begin{equation*}
    (b,c)A =
    p^\delta(va_{11}p^{\beta-\delta}+wa_{21}p^{\gamma-\delta},
    va_{12}p^{\beta-\delta}+w       a_{22}p^{\gamma-\delta})
\end{equation*}
that the degree of $(b,c)A$ is $\geq\delta$. Similarly,
$\big((b,c)A\big)A^{-1}=(b,c)$ implies that the degree of $(b,c)A$ is
$\leq\delta$. This completes the proof of (a).
\par
In order to establish (b) we distinguish two cases: If $\delta=\beta\leq\gamma$
then we put
\begin{equation*}\label{}
    M:=
    \begin{pmatrix}
    v^{-1} & -w p^{\gamma-\beta}\\0 & v
    \end{pmatrix},
\end{equation*}
whereas for $\delta=\gamma\leq\beta$ we put
\begin{equation*}\label{}
    M:=
    \begin{pmatrix}
    0 & -w \\w^{-1} & v p^{\beta-\gamma}
    \end{pmatrix}.
\end{equation*}
In either case we have $\det M=1$ and $(b,c)M=(p^\delta,0)$, as required.
\end{proof}
We add for the sake of completeness that $M^{-1}$ equals
\begin{equation*}\label{}
\begin{pmatrix}
    v & w p^{\gamma-\beta}\\0 & v^{-1}
    \end{pmatrix}
    \mbox{~~and~~}
    \begin{pmatrix}
    v p^{\beta-\gamma} & w \\-w^{-1} & 0
    \end{pmatrix},
\end{equation*}
respectively. Also, we emphasise the particular case of a vector $(b,c)$ with
degree $0$ or, said differently, of an admissible vector. Such a vector can be
moved to $(1,0)$ by an appropriate invertible matrix. This reflects the well
known fact that all points of the projective line $\bP_1(\bZ_d)$ form an orbit
under the action of the group $\GL_2(\bZ_d)$.

\begin{lem}\label{lem:invariant}
The symplectic form $[\cdot,\cdot]$ remains invariant, to within invertible
elements of $\bZ_d$, under the natural action of the general linear group
$\GL_2(\bZ_d)$ on $\bZ_d^2$ .
\end{lem}
\begin{proof}
From (\ref{eq:determinant}) follows for all $A\in\GL_2(\bZ_d)$ and all $(b,c),
(b',c')\in\bZ_d^2$ that
\begin{equation*}\label{}
    [(b,c)A,(b',c')A] = \det A [(b,c),(b',c')].
\end{equation*}
\end{proof}
This implies, in particular, that our (symplectic) orthogonality of vectors is
preserved under the natural action of $\GL_2(\bZ_d)$. We shall use Lemmas
\ref{lem:transitive} and \ref{lem:invariant} in the subsequent proofs in order
to simplify some (otherwise lengthy) calculations.

\par

\begin{thm}\label{thm:points}
Let the integer $d=p^\eps>1$ be a power of a prime $p$. Also, let $(b,c)$ be a
vector of $\bZ_d^2$ with degree $\delta$. Then the number of points of the
projective line $\bP_1(\bZ_d)$ which contain the vector $(b,c)$ equals
\begin{eqnarray*}\label{eq:pointnumber}
    p^\eps + p^{\eps-1} &\mbox{if} & \delta=\eps, \\
    p^\delta            &\mbox{if} & \delta<\eps.
\end{eqnarray*}
\end{thm}

\begin{proof}
Due to Lemmas \ref{lem:transitive} and \ref{lem:invariant}, we may confine
ourselves to the case $(b,c)=(p^\delta,0)$. Each point of $\bP_1(\bZ_d)$ can be
written in a unique way either as $\bZ_d(1,y)$, where $y\in\bZ_d$ is arbitrary,
or as $\bZ_d(x,1)$, with $x\in\bZ_d\cdot p$ (see, e.\,g., [5, page 792]). We distinguish two cases:
\par
Case 1: We have $(p^\delta,0)\in\bZ_d(1,y)$ if, and only if, $p^\delta y=0$,
which in turn is equivalent to saying that $y\in\bZ_d$ is in the annihilator of
$p^\delta$, viz. $y$ is one of the elements
\begin{equation}\label{eq:solutions}
    t p^{\eps-\delta}\in\bZ_d
    \mbox{~~with~~}t\in\{1,2,\ldots,p^\delta\}.
\end{equation}
These elements give rise to $p^\delta$ mutually distinct points containing the
vector $(p^\delta,0)$.
\par
Case 2: A point of the form $\bZ_d(x,1)$ contains the vector $(p^\delta,0)$
precisely when $(p^\delta,0)=0(x,1)=(0,0)$. Hence for $\delta<\eps$ no such
points exists, whereas for $\delta=\eps$ there are $p^{\eps-1}$ points of this
kind.
\end{proof}

Our next aim is to count the number of vectors in the perp-set of a vector
$(b,c)$.

\begin{thm}\label{thm:cardinality}
Let the integer $d=p^\eps>1$ be a power of a prime $p$. Also, let $(b,c)$ be a
vector of $\bZ_d^2$ with degree $\delta$. Then
\begin{equation*}\label{eq:cardinality}
    |(b,c)^\perp| = p^{\eps + \delta}.
\end{equation*}

\end{thm}
\begin{proof}
Again, we may assume without loss of generality that $(b,c)=(p^\delta,0)$. An
unknown vector $(x,y)\in\bZ_d^2$ belongs to $(p^\delta,0)^\perp$ if, and only
if,
\begin{equation*}\label{eq:det}
             \det \begin{pmatrix}p^\delta&0\\x &y\end{pmatrix}
             =
    p^\delta \det \begin{pmatrix} 1&0\\x &y\end{pmatrix}
    =0.
\end{equation*}
By expanding this determinant, we deduce the equivalent condition
\begin{equation}\label{eq:y-solutions}
    y \in
    \{tp^{\eps-\delta} : t=1,2,\ldots,p^\delta\}
\end{equation}
in which the unknown $x$ does not appear. So there are precisely $p^{\eps}$
solutions for $x$ and precisely $p^{\delta}$ solutions for $y$.
\end{proof}

Let us compare the results from Theorems \ref{thm:points} and
\ref{thm:cardinality}. For the sake of completeness, the following result
includes some previous findings:

\begin{thm}\label{thm:U=perp}
Let the integer $d=p^\eps>1$ be a power of a prime $p$. Also, let $(b,c)$ be a
vector of $\bZ_d^2$ with degree $\delta$. We denote by $U(b,c)\subset\bZ_d^2$
the set-theoretic union of all points of the projective line $\bP_1(\bZ_d)$
containing the vector $(b,c)$. Then $U(b,c)$ is a generating set for the
submodule $(b,c)^\perp\subset\bZ_d^2$. Furthermore, the equality
\begin{equation}\label{eq:U=perp}
    U(b,c) = (b,c)^\perp
\end{equation}
holds if, and only if, one of the following conditions is satisfied:
\begin{enumerate}
\item $(b,c)=(0,0)$.
\item $(b,c)$ is an admissible pair.
\end{enumerate}
\end{thm}

\begin{proof}
The assertions holds trivially for $(b,c)=(0,0)$, and we rule out this case for
the rest of the proof. As before, it will be assumed that $(b,c)=(p^\delta,0)$
is satisfied. We infer from (\ref{eq:y-solutions}) that $(p^\delta,0)^\perp$
equals the set of vectors of the form
\begin{equation}\label{eq:perp-vectors}
    (s,t p^{\eps-\delta})
    \mbox{~~with~~}s\in\bZ_d
    \mbox{~~and~~}t\in\{1,2,\ldots,p^\delta\}.
\end{equation}
By (\ref{eq:solutions}), a vector is in $U(p^\delta,0)$ if, and only if, it can
be written as
\begin{equation}\label{eq:U-vectors}
    (\tilde{s}, \tilde{s}\tilde{t} p^{\eps-\delta})
    \mbox{~~with~~}\tilde{s}\in\bZ_d
    \mbox{~~and~~}\tilde{t}\in\{1,2,\ldots,p^\delta\}.
\end{equation}
We have $U(p^\delta,0)\subset (p^\delta,0)^\perp$, since any vector from
(\ref{eq:U-vectors}) appears also in (\ref{eq:perp-vectors}) for $s:=\tilde s$
and the unique element $t\in\{1,2,\ldots,p^\delta\}$ which satisfies $t\equiv
\tilde s\tilde t$ (mod $p^\delta$). Conversely, each vector of
$(p^\delta,0)^\perp$ is a linear combination of vectors of $U(p^\delta,0)$,
because
\begin{eqnarray*}
    (s, t p^{\eps-\delta}) = (s-t)(1,0) + t (1,p^{\eps-\delta}),
\end{eqnarray*}
where we use on the right hand side those vectors which arise in
(\ref{eq:U-vectors}) for $(\tilde s,\tilde t):=(1,p^\delta)$ and $(\tilde
s,\tilde t):=(1,1)$. Thus $U(p^\delta,0)$ generates $(p^\delta,0)^\perp$.
\par
We infer from Theorem \ref{thm:cardinality} that equation (\ref{eq:U=perp}) is
satisfied precisely when $|U(p^\delta,0)|=p^{\eps+\delta}$. This in turn is
true if, and only if, distinct pairs $(\tilde s,\tilde t)$ determine distinct
vectors in (\ref{eq:U-vectors}). Clearly, distinct values for $\tilde s$ yield
distinct vectors, but for a fixed $\tilde s$ and a variable $\tilde t$ this
need no longer be true. Indeed, let us fix some $\tilde s\in\bZ_d$.
Furthermore, we assume that
\begin{equation}\label{eq:s-tilde}
    \tilde s=up^\sigma
\end{equation}
is a factorisation of $\tilde s$ as in (\ref{eq:power-rep}), so that $u$ is an
invertible element. For this particular value of $\tilde s$ the second
coordinate of the vector given in (\ref{eq:U-vectors}) equals
\begin{equation}\label{eq:secondcoo}
    \tilde t u p^{\eps-\delta+\sigma}.
\end{equation}
There are two cases as $\tilde t$ varies in $\{1,2,\ldots,p^{\delta} \}$:
\par
$\sigma\leq\delta$: Here (\ref{eq:secondcoo}) assumes the mutually distinct
values
\begin{equation}\label{eq:secondcoo-values}
    up^{\eps-\delta+\sigma}, 2up^{\eps-\delta+\sigma},\ldots,(p^{\delta-\sigma}-1) up^{\eps-\delta+\sigma},
    p^{\delta-\sigma} up^{\eps-\delta+\sigma}=0
\end{equation}
for $\tilde t=1,2,\ldots,p^{\delta-\sigma}$, and remains $0$ for all $\tilde
t>p^{\delta-\sigma}$.
\par
$\sigma>\delta$: The second summand is zero for all $\tilde t$.
\par
We now assume that condition (b) from the Theorem is satisfied. So the element
$p^\delta$ is invertible. This means $\delta=0$. Consequently,
$p^{\eps-\delta}=p^\eps=0$. Hence (\ref{eq:perp-vectors}) and
(\ref{eq:U-vectors}) yield the same set of $p^\eps$ vectors.
\par
Finally, assume that (b) is not satisfied. (We did already rule out (a) at the
beginning of the proof.) Thus $p^\delta$ is not invertible. This implies
$1\leq\delta$. We even have $1\leq\delta<\eps$, since $\delta=\eps$ would give
the contradiction $(p^\delta,0)=(0,0)$. By (\ref{eq:perp-vectors}), the set
$(p^\delta,0)^\perp$ has $p^\delta$ vectors of the form $(p,*)$, whereas
(\ref{eq:secondcoo-values}) shows that set $U(p^\delta,0)$ contains only
$p^{\delta-1}$ such vectors. Therefore $U(p^\delta,0)$ cannot be equal to
$(p^\delta,0)^\perp$.
\end{proof}

\par
As an appendix to the previous proof we give an explicit example of a vector
$(b,c)$ with the property $U(b,c)\neq (b,c)^\perp$. Let $d:=4$, i.~e. $p=2$ and
$\eps=2$. We exhibit the vector $(2,0)\in\bZ_4^2$. In terms of the notation
used in Theorem \ref{thm:points} we have $\delta=1\in\bZ$. There are just two
points containing $(2,0)$: These are $\bZ_4(1,0)$ and $\bZ_4(1,2)$. On the
other hand, the vector $(2,2)$ belongs to the perp-set of $(2,0)$, but it is
neither a multiple of $(1,0)$ nor of $(1,2)$.
\par
\begin{thm}\label{thm:U(b,c)-size}
Under the assumptions of Theorem \emph{\ref{thm:U=perp}} let $(b,c)$ be a
non-zero vector. Then the number of vectors of the set $U(b,c)$ equals
\begin{equation}\label{eq:U(b,c)-size}
     \Big(\sum_{\sigma=0}^{\delta-1} (p^{\eps -
     \sigma}-p^{\eps-\sigma-1})p^{\delta-\sigma}\Big)
    +  p^{\eps - \delta}.
\end{equation}
\end{thm}

\begin{proof}
We simplify matters as before by assuming that $(b,c)=(p^\delta,0)$, where
$\delta<\eps$. Choose an integer $\sigma\in\{0,1,\ldots,\eps\}$. We determine
the number of all $\tilde s=up^\sigma\in\bZ_d$, where $u$ is any invertible
element of $\bZ_d$. Observe that in contrast to (\ref{eq:s-tilde}) now only
$\sigma$ is fixed, but $\tilde s$ is variable. For $\sigma\leq\eps-1$ holds
\begin{equation*}
    \tilde s\in \bZ_d\cdot p^\sigma \setminus \bZ_d\cdot p^{\sigma+1}.
\end{equation*}
So $\tilde s$ can be chosen in $p^{\eps-\sigma}-p^{\eps-\sigma-1}$ different
ways. For $\sigma=\eps$ there is a unique choice for $\tilde s$, namely $\tilde
s=0$.
\par
Now we select one such $\tilde s$. We count how many distinct vectors arise
from (\ref{eq:U-vectors}), as $\tilde t$ varies from $1$ to $p^\delta$. By
(\ref{eq:secondcoo-values}) and the subsequent remark on the case
$\sigma>\delta$, this number of vectors equals
\begin{eqnarray*}
    p^{\delta-\sigma} &\mbox{if}& 0\leq \sigma\leq\delta-1,  \nonumber\\
    1                &\mbox{if}& \delta\leq \sigma\leq\eps.
\end{eqnarray*}
Note that result depends only on $\sigma$, but not on $\tilde s$.
\par
Finally, we regard $\sigma$, $\tilde s$, $\tilde t$ to be variable and count
the maximal number of pairs $(\tilde s,\tilde t)$ which give rise to distinct
vectors in (\ref{eq:U-vectors}). As $\sigma$ ranges from $0$ to $\delta-1$, the
maximal number of such pairs is given by the sum on the left hand side of
(\ref{eq:U(b,c)-size}). For $\delta<\sigma\leq\eps$ we obtain
\begin{equation*}\label{}
    \sum_{\sigma=\delta}^{\eps-1}(p^{\eps-\sigma}-p^{\eps-\sigma-1}) + 1 =
    p^{\eps-\delta}
\end{equation*}
such pairs. This completes the proof.
\end{proof}
Note that (\ref{eq:U(b,c)-size}) remains meaningful for $(b,c)=(0,0)$, but it
\emph{does not provide the correct number of vectors\/} for
$(0,0)^\perp=\bZ_d^2$. This is due to the fact that in (\ref{eq:U-vectors}) we
disregard those points which appear (for $(b,c)=(0,0)$ only) in the proof of
Theorem \ref{thm:points}, Case 2.

\section{The case of an arbitrary qudit}

Throughout this section we adopt the assumption that
\begin{equation}\label{eq:factors}
    d = p_1^{\eps_1}p_2^{\eps_2}\cdots p_r^{\eps_r},
\end{equation}
where $p_1,p_2,\ldots,p_r$ are $r\geq 1$ distinct prime numbers, and the
exponents $\eps_1,\eps_2,\ldots,\eps_r$ are non-negative integers $\geq 1$.
Furthermore, we let
\begin{equation*}\label{}
    d_k:=p_k^{\eps_k} \mbox{~~for all~~}k\in\{ 1,2,\ldots,r \}.
\end{equation*}
It is well known that the ring $(\bZ_d,+,\cdot)$ is isomorphic to the outer
direct product
\begin{equation}\label{eq:outer}
    \bZ_{d_1}\times\bZ_{d_2}\times\cdots\times\bZ_{d_r}.
\end{equation}
An isomorphism is given by assigning to each $x\in\bZ_d$ the $r$-tuple
\begin{equation*}
    (x^{(1)},x^{(2)},\ldots,x^{(r)})\in
    \bZ_{d_1}\times\bZ_{d_2}\times\cdots\times\bZ_{d_r},
\end{equation*}
where $x^{(k)} \equiv x \pmod {d_k}$ for all $k\in\{1,2,\ldots,r \}$. We use
this isomorphism to identify $\bZ_d$ with the outer direct product given in
(\ref{eq:outer}), i.~e., we do not distinguish between $x\in\bZ_d$ and the
$r$-tuple of its \emph{components} $x^{(k)}$. Addition and multiplication of
these $r$-tuples is carried out componentwise, and calculations in the $k$th
component are understood modulo $d_k$. Note that we used a representation of
$\bZ_d$ as the \emph{inner direct product\/} of $r$ ideals in \cite{hs}. In the
present paper we shall not follow that approach.
\par
This representation of $\bZ_d$ as a direct product has several straightforward
consequences for the $\bZ_d$-module $\bZ_d^2$: Given a vector $(b,c)\in\bZ_d^2$
we define its \emph{component vectors} as $(b^{(k)},c^{(k)})\in\bZ_{d_k}^2$ for
all $k\in\{1,2,\ldots,r \}$. The \emph{degree\/} of $(b,c)\in\bZ_d^2$ is that
$r$-tuple
\begin{equation}\label{eq:delt}
    \delta:=(\delta_1,\delta_2,\ldots,\delta_r)
\end{equation}
which is formed by the degrees of its component vectors (in natural order).
Thus, for example, the zero-vector of $\bZ_d$ is the only vector with degree
$(\eps_1,\eps_2,\ldots,\eps_r)$. A vector $(b,c)\in\bZ_d$ is admissible if, and
only if, there exist elements $u,v\in\bZ_d$ with
\begin{equation*}
    u^{(k)}b^{(k)}+v^{(k)}c^{(k)}=1\mbox{~~for all~~}
    k\in\{1,2,\ldots,r\}.
\end{equation*}
This is equivalent to saying that all component vectors of $(b,c)$ are
admissible which in turn means that the degree of $(b,c)$ equals
$(0,0,\ldots,0)$.
\par
Likewise, each submodule of $\bZ_d^2$ can be split into its components. The
following important observation is immediate from the above: A submodule of
$\bZ_d^2$ is free and cyclic (i.~e. a point) if, and only if, all its
components are free and cyclic. Thus the projective line over $\bZ_d$ can be
viewed as the Cartesian product
\begin{equation}\label{eq:lineproduct}
    \bP_1(\bZ_{d_1})\times\bP_1(\bZ_{d_2})\times\cdots\times\bP_1(\bZ_{d_r}).
\end{equation}
This allows to carry over the results from Section \ref{sect:primepower} to our
more general setting.
\par
Of course, also each matrix $A\in\GL_2(\bZ_d)$ can be split into its
\emph{component matrices} $A^{(k)}\in\GL_2(\bZ_{d_k})$. Lemma
\ref{lem:transitive} implies that the degree of vectors of $\bZ_d^2$ is a
$\GL_2(\bZ_{d})$-invariant notion. Furthermore, each vector with degree
$\delta=(\delta_1,\delta_2,\ldots,\delta_r)$ can be mapped to a vector $(q,0)\in\bZ_d^2$ with
$q^{(k)}=p_k^{\delta_k}$ for all $k\in\{1,2,\ldots,r\}$. Likewise, Lemma
\ref{lem:invariant} holds, {\it mutatis mutandis}, for an arbitrary $d$.
\par
We are now in a position to extend our Theorems \ref{thm:points} and
\ref{thm:cardinality}.

\begin{thm}\label{thm:d-points+card}
Let the integer $d$ be given as in \emph{(\ref{eq:factors})}. Also, let $(b,c)$
be a vector of $\bZ_d^2$ with degree
$\delta=(\delta_1,\delta_2,\ldots,\delta_r)$. We denote by $K$ the set of those
indices $k\in\{1,2,\ldots,r\}$ such that $(b^{(k)},c^{(k)})=(0,0)$. Then the
following assertions hold:\footnote{Below we use the shorthand $j\notin K$ for
$j\in\{1,2,\ldots,r\}\setminus K$.}
\begin{enumerate}
\item The number of points of the projective line $\bP_1(\bZ_d)$ which contain
the vector $(b,c)$ equals
\begin{equation*}\label{eq:pointnumber-d}
    \prod_{j\notin K} p_j^{\delta_j}
    \cdot
    \prod_{k\in K}(p_k^{\eps_k} + p_k^{\eps_k-1}).
\end{equation*}
\item
The perp-set $(b,c)^\perp$ has cardinality
\begin{equation*}\label{eq:perp-gen}
    |(b,c)^\perp| = \prod_{k=1}^{r}p_{k}^{\eps_k+\delta_k} =
    d\cdot\prod_{k=1}^{r}p_{k}^{\delta_k}.
\end{equation*}
\end{enumerate}
\end{thm}
\begin{proof}
It suffices to apply Theorem \ref{thm:points} and Theorem \ref{thm:cardinality}
to the component vectors of $(b,c)$ and to multiply the cardinalities which can
be read off from there.
\end{proof}
\noindent
Since each pair $(b,c)$ corresponds to all operators of the form $\omega^a X^b Z^c$, and there are $d$ such operators, as an important corollary
we have
\begin{cor}
The number of operators in the generalized Pauli group $G$ which
commute with the operator $\omega^a X^bZ^c\in G$ is equal to
\begin{equation}\label{eq:oper}
   d \cdot |(b,c)^\perp| = d^2\cdot\prod_{k=1}^{r}p_{k}^{\delta_k},
\end{equation}
where $(\delta_1,\delta_2,\ldots,\delta_r)$ is the degree of $(b,c)$. 
\end{cor}

We may define $U(b,c)$ just in the same way as in Section \ref{sect:primepower}
as the set-theoretic union of all points of the projective line $\bP_1(\bZ_d)$
which contain the vector $(b,c)$. By our identification of $\bP_1(\bZ_d)$ with
the Cartesian product (\ref{eq:lineproduct}), it is immediately clear that
Theorem \ref{thm:U=perp} holds, {\it mutatis mutandis}, also for our arbitrary $d$.
\par
Our last result in this section is the following straightforward generalisation
of Theorem \ref{thm:U(b,c)-size}:

\begin{thm}\label{thm:d-U(b,c)-size}
Under the assumptions of Theorem~\emph{\ref{thm:d-points+card}} let $(b,c)$ be
a non-zero vector. Then the number of vectors of the set $U(b,c)$ equals
\begin{equation*}\label{eq:d-U(b,c)-size}
     \prod_{j\notin K}\left(
     \Big(\sum_{\sigma_j=0}^{\delta_j-1} (p_j^{\eps_j -
     \sigma_j}-p_j^{\eps_j-\sigma_j-1})p_j^{\delta_j-\sigma_j}\Big)
    +  p_j^{\eps_j - \delta_j}
    \right)\cdot
    \prod_{k\in K}d_k^{2}.
\end{equation*}
\end{thm}
\begin{proof}

For all $j\notin K$ we can apply Theorem~\ref{thm:d-points+card} in order to
obtain the number of vectors in the $j$th component of $U(b,c)$. For the
remaining indices $k\in K$ the $k$th component of $U(b,c)$ coincides with
$\bZ_{d_k}^2$, and this is a set with cardinality $d_k^2$. The proof is now
accomplished by multiplying these numbers.
\end{proof}

\section{Discussion and conclusion}
A detailed study of a single qudit living in the Hilbert space of
an arbitrary finite dimension $d$ has been performed in terms of
the commutation algebra of the elements of the corresponding
generalized Pauli group $G$. The principal outcome of this
analysis is the universal formula for the number of operators
commuting with a given one (eq.\,(\ref{eq:oper})) and its
interpretation in terms of the fine structure of the projective
line defined over the modular ring $\bZ_{d}$. As each operator of
the group $G/G'$ has the unique counterpart in a vector of
$\bZ_{d}^{2}$, it belongs to a certain `layer' characterized by
the degree $\delta$ of the corresponding vector (see
eq.\,(\ref{eq:delt})). In light of eq.\,(2), the whole set of the
generalized Pauli operators is thus naturally structured into
disjoint layers. The uppermost layer, $\delta=(0, 0,\ldots, 0)$,
comprises all those operators which correspond to admissible
vectors, while all the remaining layers feature operators
represented by non-admissible vectors; the lowermost layer,
$\delta = (\eps_1, \eps_2,\ldots, \eps_r)$, consisting of $d$
operators of $Z(G)$ (eq.\,(5)).\footnote{Very roughly speaking, the
greater the value of $\Delta \equiv \delta_1 + \delta_2 + \cdots +
\delta_r$, the `lower' the layer; this and some other novel, and
rather unexpected, properties of the structure of finite
projective ring lines deserve a careful treatment of their own and
will, therefore, be the subject of a separate paper.} Given the
fact that the value of $\delta$ is intimately connected with the
number of free cyclic submodules of $\bP_1(\bZ_d)$ shared by a
given vector, this layered structure of the operators' set can be
given a nice geometrical representation, as illustrated in Fig.\,1
for $d=12$ (i.\,e., for $p_1 = 3, p_2 = 2, \eps_1 = 1 ~{\rm and}~
\eps_2 = 2$). 
\begin{figure}[pht!]
\centerline{\includegraphics[width=10.0cm,clip=]{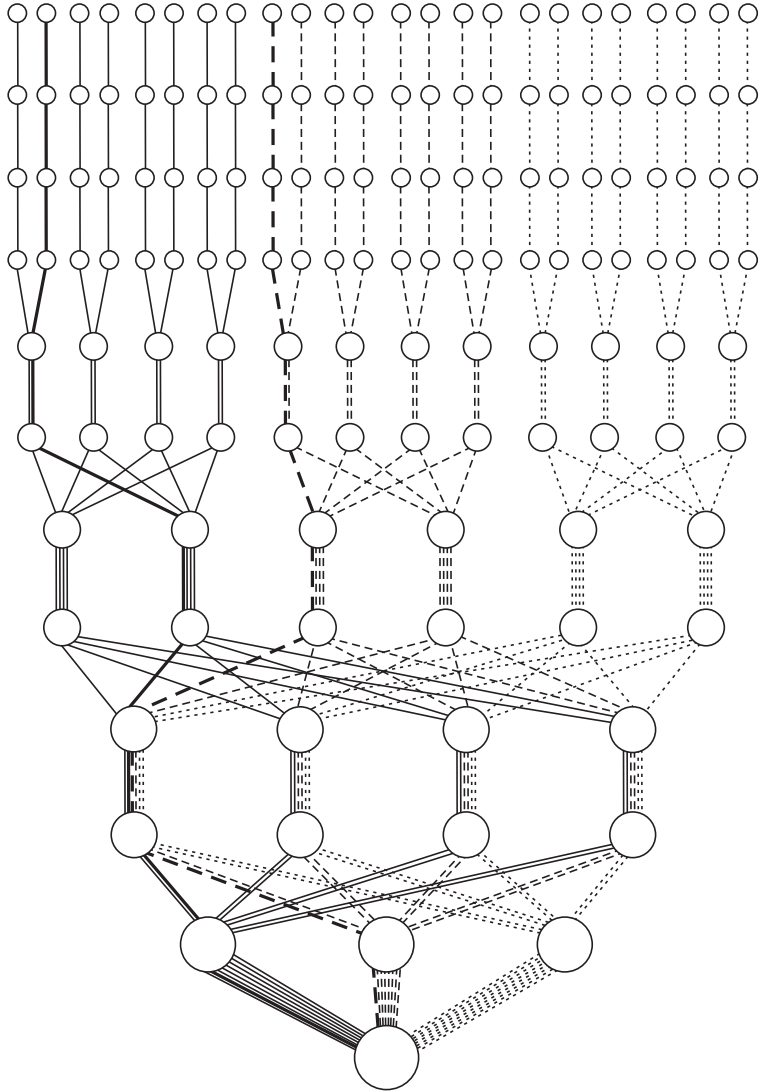}}
\vspace*{.2cm} \caption{A schematic illustration of the layered
structure of the set of the generalized Pauli operators of a
12-dimensional single qudit. Each circle represents $d$ operators
$\omega^a X^b Z^c$ with $b$ and $c$ fixed, that is, one vector of
$\bZ_{12}^{2}$, and its size increases with the increasing number
of free cyclic submodules `passing' through it. The circles are
arranged into twelve horizontal rows; the four rows at the top
characterize admissible vectors, the one at the very bottom
accommodating all the operators of $Z(G)$. Each free cyclic
submodule consists of twelve circles, one from each row, joined by
line segments in an obvious way; a couple of them are boldfaced so
that one can readily recognize a generic shape. A layer is created
by the circles of the same size. It can easily be discerned that
this particular qudit features six layers characterized (top to
bottom) by the following values of $\delta=(\delta_1,\delta_2)$: (0,0), (0,1), (1,0),
(0, 2), (1,1) and (1,2) and having the following cardinalities (in multiples of 12):
96, 24, 12, 8, 3 and 1, respectively. (Three different kinds of
shading of line segments are used only to make the case more
illustrative.)}
\end{figure}
This representation acquires an especially remarkable form 
when $d$ is a product of distinct primes \cite{hs}, i.\,e., when  $\eps_1 
= \eps_2 = \ldots = \eps_r = 1$ ($r>1$). In this case $r$-tuples 
$(\delta_1, \delta_2,\ldots, \delta_r)$ of {\it non}-admissible vectors can 
be regarded as coordinates of the points of the $(r-1)$-dimensional projective space 
over $\bZ_2$, $PG(r-1,2)$, which means that each `non-admissible' {\it layer} of 
the generalized Pauli operators corresponds to a {\it point} of $PG(r-1,2)$; note that this 
correspondence is sensitive only to the number of factors in (\ref{eq:factors}), not
to the values of the factors themselves.
\par
From a physical point of view, it is interesting to mention the maximal
sets of mutually perpendicular vectors {\it aka} the maximal sets of mutually commuting
Pauli operators. Obviously, any free cyclic submodule $\bZ_d(b,c)$ with ($b,c$) admissible 
is such a set. Yet, there are also others; for example, over $\bZ_4$ we find the set 
$\{(0,0), (2,2), (2,0), (0,2) \}$ that is not a free cyclic submodule. So the question
of the properties and cardinalities of such sets for a generic case is a challenging open problem.
\par
Next, an interesting technical aspect of our approach should be emphasized.
The attentive reader might have noticed that although we started with vectors
of $\bZ_d^2$ and their symplectic module, we finally (Theorem \ref{thm:U=perp}) reformulated the problem 
of finding the perp-set $(b,c)^\perp$ in purely geometrical terms, just employing the points of the projective line
$\bP_1(\bZ_d)$ containing $(b,c)$ and their span. This is, however, in marked contrast to how multiple-qudit
cases are handled \cite{th1,th2} for there symplectic form is essential --- even in the simplest, multiple-qubit cases \cite{adv}. 
This means that although
for our {\it single} qudits the [.,.] form seems to be a redundant concept, it is expected to play a crucial role when
extending this approach to tensorial-qudit cases.
\par
As a concluding remark, we would like to stress that it is the above-discussed 
layered structure of generalized Pauli operators
which, in our opinion, is a major feature distinguishing a single $d$-qudit from a
`tensorial' multi-qudit of the same dimension. Here, the $d=4$
case can serve as an elementary illustration of this fact; while
our single 4-qudit is characterized by two non-trivial layers
(disregarding the trivial $Z(G)$-layer) which are
embodied in the structure of the projective line over $\bZ_{4}$, a
two-qubit features just a single layer since the geometry behind
the corresponding tensor products of the classical Pauli matrices
is that of the generalized quadrangle of order two
\cite{spp}--\cite{qic}.  Similar comparisons can also be made for
several other low-dimensional quantum systems
\cite{adv,pbs,pb}. These should prove helpful when extending
this group-geometrical approach to the most general case of
multiple qudits.

\section*{Acknowledgements}
This work was supported by the Science and Technology Assistance
Agency under the contract $\#$ APVT--51--012704, the VEGA grant
agency projects $\#$ 2/6070/26 and $\#$ 7012 and by the
$\diamond$Action Austria--Slovakia$\diamond$ project $\#$ 58s2
``Finite Geometries Behind Hilbert Spaces." We thank Dr. Petr
Pracna for creating an electronic version of the figure.

\end{document}